\documentclass[aps,preprint,superscriptaddress]{revtex4-1}
\usepackage{amssymb}
\usepackage[T1]{fontenc}
\usepackage[latin9]{inputenc}
\usepackage{mathrsfs}
\usepackage{amsmath}
\usepackage{amsfonts}
\usepackage{graphicx}
\usepackage{esint}
\usepackage[unicode=true,pdfusetitle,
 bookmarks=true,bookmarksnumbered=false,bookmarksopen=false,
 breaklinks=false,pdfborder={0 0 1},backref=false,colorlinks=false]{hyperref}
\usepackage{breakurl}
\usepackage{fixmath}
\usepackage{wrapfig}
\usepackage[toc,page]{appendix}
\usepackage{mathrsfs}
\usepackage{setspace}
\usepackage{breakurl}

\makeatletter
\graphicspath{{images/}}

\newcommand{\be}{\begin{equation}}
\newcommand{\ee}{\end{equation}}
\makeatother

\begin{document}

\title{Inverse Faraday effect in graphene and Weyl semimetals}
\author{I.D.Tokman }
\affiliation{Institute for Physics of Microstructures, Russian Academy of Sciences, Nizhny Novgorod, 603950, Russia }
\affiliation{Lobachevsky State University of Nizhny Novgorod, Nizhny Novgorod, 603950, Russia}
\author{ Qianfan Chen}
\affiliation{Department of Physics and Astronomy, Texas A\&M University,
College Station, TX, 77843 USA}
\author{I.A. Shereshevsky}
\affiliation{Institute for Physics of Microstructures, Russian Academy of Sciences, Nizhny Novgorod, 603950, Russia }
\affiliation{Lobachevsky State University of Nizhny Novgorod, Nizhny Novgorod, 603950, Russia}
\affiliation{Sirius University of Science and Technology, 354340 Sochi, Russia}
\author{V.I.Pozdnyakova}
\affiliation{Institute for Physics of Microstructures, Russian Academy of Sciences, Nizhny Novgorod, 603950, Russia }
\affiliation{Lobachevsky State University of Nizhny Novgorod, Nizhny Novgorod, 603950, Russia}
\affiliation{Sirius University of Science and Technology, 354340 Sochi, Russia}
\author{Ivan Oladyshkin}
\affiliation{Institute of Applied Physics, Russian Academy of Sciences, Nizhny Novgorod, 603950, Russia }
\author{ Mikhail Tokman}
\affiliation{Institute of Applied Physics, Russian Academy of Sciences, Nizhny Novgorod, 603950, Russia }
\author{ Alexey Belyanin}
\affiliation{Department of Physics and Astronomy, Texas A\&M University,
College Station, TX, 77843 USA}

\begin{abstract}
We report systematic theoretical studies of the inverse Faraday effect in materials with massless Dirac fermions, both in two dimensions such as graphene and surface states in topological insulators, and in three dimensions such as Dirac and Weyl semimetals.  Both semiclassical and quantum theories are presented, with dissipation and finite size effects included. We find that the magnitude of the effect can be much stronger in Dirac materials as compared to conventional semiconductors. Analytic expressions for the optically induced magnetization in the low temperature limit are obtained. Strong inverse Faraday effect in Dirac materials can be used for the optical control of magnetization, all-optical modulation, and optical isolation. 
\end{abstract}

\date{ \today }
\maketitle

\section{Introduction}

Inverse Faraday Effect (IFE) is a fascinating nonlinear optical phenomenon.
Its key feature is generation of a permanent magnetization in a medium as a
result of interaction with circularly polarized radiation \cite{landau1984}.
The effect was predicted by Pitaevskii \cite{pitaevskii1960}, and the
name IFE was coined in \cite{ziel1965,pershan1963,pershan1966}. IFE was studied extensively in plasmas,
metals, and semiconductors \cite{karpman1982, mdtokman 1984, horovitz1997,
najmudin2001, nasery2010, hertel2006, riccardo2015, idtokman1999}. More
recent studies explored the use of IFE for ultrafast modulation of
magnetization with femtosecond laser pulses \cite%
{stanciu2007,vahaplar2012,hansteen2006,reid2010prl,makino2012,jin2011,satoh2010,reid2010prb,kirilyuk2010,iida2011}%
.

There has been a lot of recent interest in the optical properties of
2D and 3D materials with Dirac and Weyl fermions, including the nonlinear
optical \cite{glazov2014, bonaccorso2010, otsuji2012, glazov2011,
mikhailov2017, yao2014,  mdtokman2016, wang2016, cheng2017,
oladyshkin2017, dean2010, smirnova2014, mikhailov2016, cheng2014,
mdtokman2019} and magnetooptical \cite{mdtokman2013, mikhailov2009,
mdtokman2014, yao2015, kutayiah2018, long2018} response of graphene and
Dirac/Weyl semimetals.  Strong light-matter coupling
in these systems makes them promising for IFE studies. In \cite{glazov2014,
karch2011} the generation of edge photocurrent in graphene was studied
theoretically and in experiments. We show below that generation of edge
photocurrent is related to IFE.

In the Introduction we discuss general features of IFE. Section II
derives a quasiclassical expression for magnetization of graphene monolayer.
The quantum-mechanical derivation including interband transitions is given
in Sec.~III. In Sec.~IV we discuss peculiarities of IFE in dissipative systems. Sec.~V takes into account finite-size effects and calculates edge photocurrent. Sec.~VI 
describes IFE in Weyl semimetals. In Appendix A we evaluate the effect of the 
depolarization field on the IFE in a finite sample, whereas Appendix B studies
saturation of IFE in strong fields.

In a transparent nonmagnetic medium, i.e. in the medium with magnetic
permeability $\mu =1$, the magnetization excited by a monochromatic field
can be determined from thermodynamic considerations. The resulting
expression is \cite{landau1984}:
\begin{equation}
\label{1}
\mathbf{m}=\sum_{ij}\frac{\partial \varepsilon _{ij}}{\partial \mathbf{H}}%
\frac{\widetilde{E}_{j}\widetilde{E}_{i}^{\ast }}{16\pi },\
\end{equation}%
where the optical field is given by $\mathbf{E}=\mathrm{Re}\left( \mathbf{%
\tilde{E}}e^{-i\omega t}\right) $, \textit{i}, \textit{j} are Cartesian
indices, $\varepsilon _{ij}$ is a Hermitian tensor of the dielectric
permittivity, $\mathbf{H}$ is the vector of a constant magnetic field. Here
the Gaussian units are assumed. In the absence of an external magnetic
field, the derivative in Eq.~(\ref{1}) should be calculated in the limit $\lim_{%
\mathbf{H}\longrightarrow 0}\left( \frac{\partial \varepsilon _{ij}}{%
\partial \mathbf{H}}\right) $ . If the medium is isotropic at $\mathbf{H}%
\rightarrow 0$ the induced magnetic moment will be orthogonal to the plane
containing the electric field vector $\mathbf{E}$ (see Fig.~1). The
magnitude of magnetization is determined by the difference between the
intensities of right- and left-circularly polarized components of the
optical field. It is obviously zero for a linearly polarized field.

\begin{figure}[htb]
\begin{center}
\includegraphics[scale=0.4]{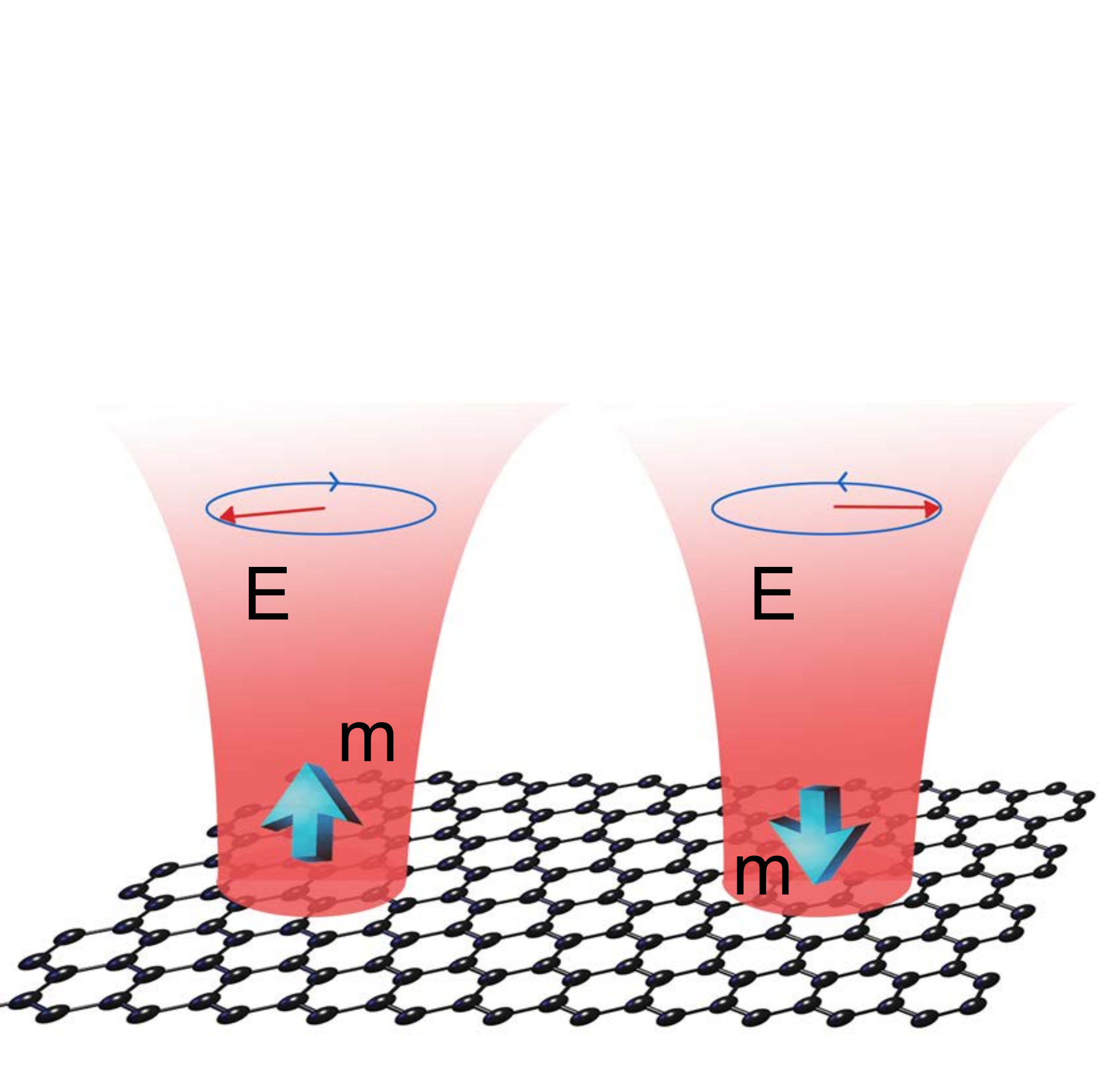}
\caption{A sketch of inverse Faraday effect: an incident circularly polarized light induces magnetization in a sample. }
\label{fig1}
\end{center}
\end{figure}

It is remarkable that Eq.~(\ref{1}) remains valid for media with frequency
dispersion: there is no need to add frequency derivatives $\frac{\partial
\varepsilon_{ij}}{\partial \omega }$ to Eq.~(\ref{1}) whereas such derivatives
are present in the expression for an averaged energy of the optical field in
a dispersive medium \cite{landau1984, pitaevskii1960}.

Equation (\ref{1}) underscores another unique feature of the IFE. It is well known
that any optical response that is quadratic in powers of the field can be
calculated within a standard perturbative approach from the second-order
(with respect to the field) perturbation of the density matrix. For a
photoinduced magnetic moment in a system with discrete energy spectrum such
an approach was developed e.g. in \ \cite{battiato2014}. At the same time,
Eq.~(\ref{1}) shows that it is possible to calculate photoinduced magnetization from the 
 \textit{linear} dielectric response of the medium.

It follows from Eq.~(\ref{1}) that IFE exists only in the media that become
gyrotropic in an external constant magnetic field. Examples of the systems
that \textit{do not} become gyrotropic in an external magnetic field include
an electron-positron plasma and magnetized vacuum \cite{ginzburg1979}.
Condensed matter systems with complete electron-hole symmetry are also not
gyrotropic in an external magnetic field. One obvious example is a material
with electronic bandstructure in the form of isotropic Dirac cones, when the
Fermi level crosses the Dirac points, such as graphene or certain types of
Dirac/Weyl semimetals \cite{long2018}. Of course this also implies low
enough photon frequencies that probe only the range of electron energies
close to the Dirac point. The selection rules for such systems allow one to
group all electric-dipole allowed optical transitions into symmetric pairs $%
n\rightarrow -\left( n+1\right) $ and $n+1\rightarrow -n$ with the same
transition frequency but opposite direction of rotation of a circularly
polarized optical field \cite{long2018, zheng2002, mcclure1956}. Gyrotropy,
and therefore the IFE, will appear in these materials only when the Fermi
level is shifted with respect to the Dirac/Weyl point; see Fig.~2. Moreover, as we argue below, the IFE is strongest in the limit of small
frequencies and large Fermi energies, when resonant interband transitions
are Pauli-blocked minimizing absorption and the main contribution to IFE
comes from intraband transitions in the vicinity of the Fermi level.

\begin{figure}[htb]
\begin{center}
\includegraphics[scale=0.4]{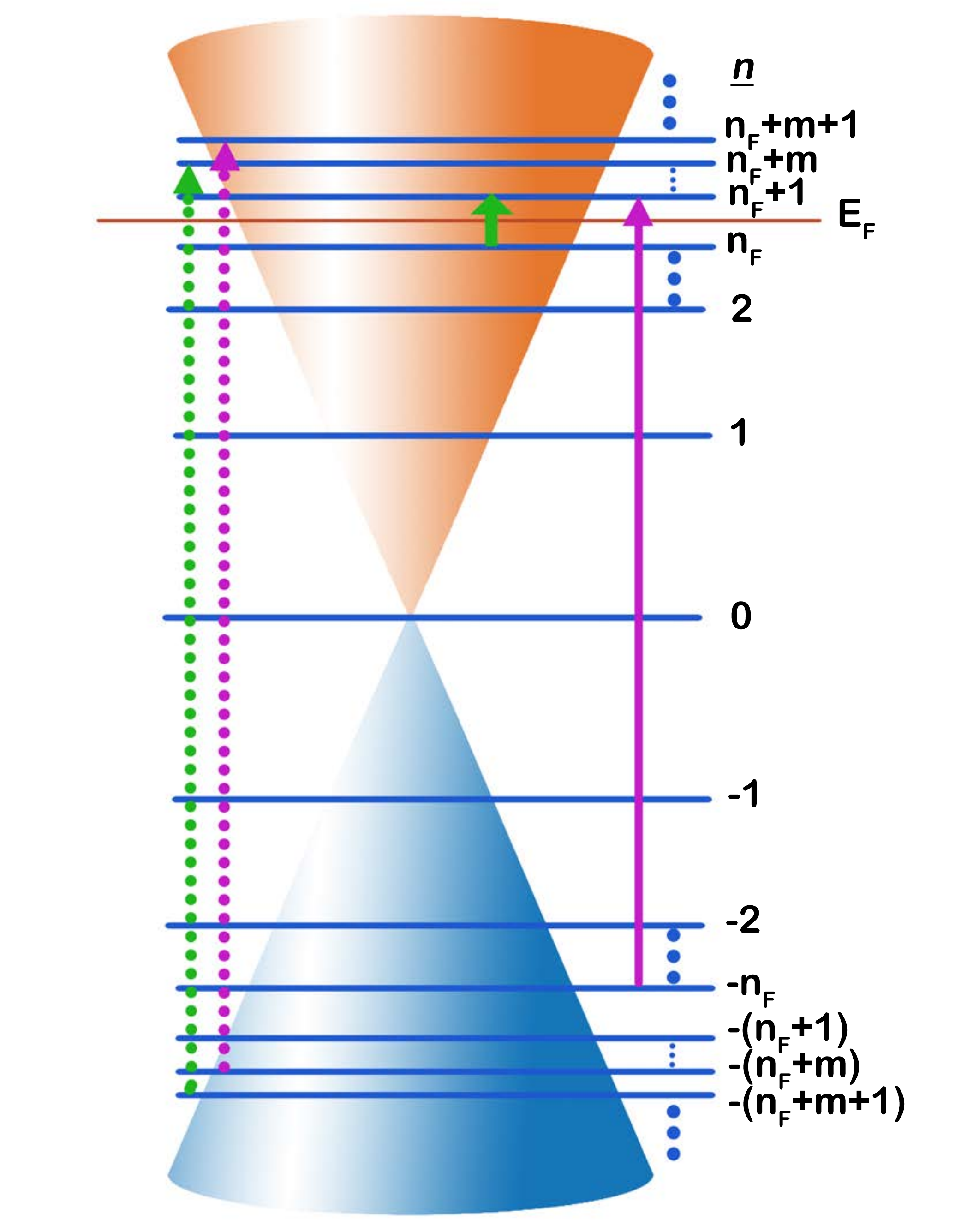}
\caption{Landau levels and optical transitions in graphene. The highest Landau level below the Fermi energy is denoted as $n_F$. Dotted arrows indicate a pair of transitions with contributions to the induced magnetic moment that cancel each other. Only the transitions shown with solid arrows (one interband and one intraband) contribute to inverse Faraday effect at low temperature. }
\label{fig2}
\end{center}
\end{figure}

Since the model leading to Eq.~(\ref{1}) does not include dissipation, for
condensed matter systems it can give only a qualitative description.
Nevertheless, it provides a useful limit based on general thermodynamic
relations\ that can be compared with a specific microscopic model that does
take dissipation into account.

\section{Quasiclassical theory of IFE in graphene}

For a 2D system such as graphene, it is convenient to use the electric
susceptibility tensor instead of the dielectric permittivity in Eq.~(1),
namely $\chi _{ij}=\frac{\varepsilon _{ij}-\delta _{ij}}{4\pi }$ , and
integrate this equation over the layer thickness. In this case Eq.~(1)
becomes 
\begin{equation}
\mathbf{m}=\sum_{ij}\frac{\partial \chi _{ij}}{\partial \mathbf{H}}\frac{\widetilde{E}%
_{j}\widetilde{E}_{i}^{\ast }}{4}.
\end{equation}%
Now the tensor $\chi _{ij}$ is a 2D surface susceptibility tensor which has
the dimension of length; $i,j=x,y$ are coordinates in the graphene plane. The vector $\mathbf{m}$ in
Eq.~(2) has a meaning of a magnetic moment of a unit area (see Fig.~1).  We will use a
standard low-energy effective Hamiltonian for electrons near the Dirac point  
\cite{katsnelson2012}:
\begin{equation}
\hat{H}_{0}=v_{F}\mathbf{\hat{p}}\cdot \boldsymbol{\hat{\sigma}},
\end{equation}%
where $\boldsymbol{\hat{\sigma}} =\mathbf{x}_{0}\hat{\sigma}_{x}+\mathbf{y}_{0}%
\hat{\sigma}_{y}$, $\mathbf{\hat{p}}=\mathbf{x}_{0}\hat{p}_{x}+\mathbf{y}_{0}%
\hat{p}_{y}$, $\hat{\sigma}_{x,y}$ are Pauli matrices, $\hat{p}_{x,y}$ are
Cartesian components of the momentum operator, $\mathbf{x}_{0}$, $\mathbf{y}%
_{0}$ are unit vectors of coordinate axes, $v_{F}$ is the Fermi velocity. In
this model the degeneracy factor $g=4$ (two spin states and two valleys).
The corresponding electron energies are%
\begin{equation}
W(p)=sv_{F}p,
\end{equation}%
where $p=\sqrt{p_{x}^{2}+p_{y}^{2}}$ ; index $s=\pm 1$ corresponds to the
conduction and valence band, respectively.

The analysis below is applicable also to 2D surface states in 3D topological insulators such as Bi$_2$Se$_3$. Their low-energy Hamiltonian is related to that of graphene by a unitary transformation, and the resulting linear and nonlinear optical responses are both very similar, after rescaling the values of the Fermi velocity and degeneracy, see e.g.~\cite{yao2014,yao2015,wang2016}. 

Since in this model the IFE appears only when the Fermi energy is shifted
from the Dirac point, we consider doped graphene and assume that the Fermi
level is in the conduction band for definiteness. In the limit of small
enough frequencies, low temperatures, and large Fermi energies (so that the
contribution of interband transitions can be neglected) the quasiclassical
theory is adequate. (This is the most interesting limit anyway: the
results for a classical plasma, metals, and semiconductors \cite%
{karpman1982, mdtokman 1984, horovitz1997, najmudin2001, nasery2010,
hertel2006, riccardo2015,idtokman1999} indicate that the photogenerated
magnetic moment grows with decreasing frequency as $\propto \omega ^{-3}$.)
Indeed, it was shown in \cite{wang2016} that under rather weak restrictions
on the nonuniformity of the electromagnetic field in the plane of graphene
both linear and quadratic intraband susceptibilities derived within the
quantum-mechanical density matrix formalism coincide with the results
obtained from the kinetic equation based on the quasiclassical equations of
motion for carriers. The nonuniformity restriction is $L\gg \frac{\hbar }{%
p_{F}}$ , where $L$ is the spatial scale of the nonuniformity of the field and $%
p_{F}$ is the Fermi momentum related to the Fermi energy by $%
W_{F}=v_{F}p_{F} $. The contribution of interband transitions will be small
when electrons are degenerate and%
\begin{equation}
W_{F}\gg \hbar \omega .
\end{equation}%
This is confirmed by fully quantum treatment in Sec. III.

Under a more restrictive condition $L\gg \frac{v_{F}}{\omega }$ one can
calculate the response neglecting spatial nonuniformity of the optical field
\cite{wang2016}. We will use the kinetic equation which corresponds to the
quasiclassical equations of motion \cite{glazov2011,mikhailov2017,
smirnova2014,mikhailov2009,mdtokman2014}. To calculate the derivative in
Eq.~(2) it is enough to know the dependence of the tensor elements $\chi
_{ij} $ on the external constant magnetic field in linear approximation with
respect to $\mathbf{H}$. Here the magnetic field is orthogonal to the
monolayer: $\mathbf{H=z}_{0}H_{z}$. The kinetic equation has the form 
\begin{equation}
\label{6}
\frac{\partial f}{\partial t}-e\left\{ \mathbf{E}\left( t\right) +[\frac{1}{c%
}\left( \frac{\partial W}{p\partial p}\right) \mathbf{p}\times \mathbf{H}%
]\right\} \cdot \frac{\partial f}{\partial \mathbf{p}}=\hat{Q}\left(
f\right) .
\end{equation}%
Here $\hat{Q}\left( f\right) $ is the relaxation operator, the electric
field vector $\mathbf{E}$ is in the graphene plane, $-e$ is electron charge.
We don't specify any particular electron dispersion $W(p)$ in Eq.~(6) in
order to compare the results for linear and quadratic dispersion (see also
\cite{idtokman1999}).

Consider Eq.~(6) when $\hat{Q}\left( f\right) $ $=0$. We need to calculate
the linear response to the uniform high-frequency field $E_{x,y}=\mathrm{Re}%
\left( \tilde{E}_{x,y}~e^{-i\omega t}\right) $. We will seek the solution to
Eq.~(6) in the form $f=\mathrm{Re}\left[ \delta f\left( \theta ,p\right)
e^{-i\omega t}\right] +f_{F}\left( p\right) ,$ where $p_{x}=p\cos \theta ,$ $%
p_{y}=p\sin \theta ,\left\vert \delta f\right\vert \ll f_{F}$. Linearization
of Eq.~(6) gives%
\begin{equation*}
-i\omega \delta f+\frac{\partial W}{p\partial p}\frac{eH_{z}}{c}\frac{%
\partial \delta f}{\partial \theta }-e\left( \tilde{E}_{x}\cos \theta +%
\tilde{E}_{y}\sin \theta \right) \frac{\partial f_{F}}{\partial p}=0.
\end{equation*}%
This equation has an exact solution:%
\begin{equation}
\delta f=\frac{e}{\omega ^{2}-\left( \frac{\partial W}{p\partial p}\frac{%
eH_{z}}{c}\right) ^{2}}\frac{\partial f_{F}}{\partial p}\left[ \tilde{E}%
_{x}\left( i\omega \cos \theta -\frac{\partial W}{p\partial p}\frac{eH_{z}}{c%
}\sin \theta \right) +\tilde{E}_{y}\left( i\omega \sin \theta +\frac{%
\partial W}{p\partial p}\frac{eH_{z}}{c}\cos \theta \right) \right] .
\end{equation}%
The surface current is determined by%
\begin{eqnarray*}
j_{x} &=&-eg\mathrm{Re}\left( e^{-i\omega t}\int \frac{\partial W}{\partial p%
}\cos \theta \delta fd^{2}p\right) , \\
j_{y} &=&-eg\mathrm{Re}\left( e^{-i\omega t}\int \frac{\partial W}{\partial p%
}\sin \theta \delta fd^{2}p\right).
\end{eqnarray*}%
Substituting Eq.~(7) in these equations and keeping only the terms linear
with respect to the magnetic field we obtain the following expressions for
the elements of the conductivity tensor $\sigma _{ij}$ :%
\begin{equation}
\begin{array}{l}
\displaystyle \sigma _{xx} =\sigma _{yy}=\sigma =-i\frac{g\pi e^{2}}{\omega }%
\int_{0}^{\infty }\frac{\partial W}{\partial p}\frac{\partial f_{F}}{%
\partial p}pdp,   \\
\displaystyle \sigma _{xy} =-\sigma _{yx}=-\frac{e^{3}g\pi H_{z}}{\omega ^{2}c}%
\int_{0}^{\infty }\left( \frac{\partial W}{\partial p}\right) ^{2}\frac{%
\partial f_{F}}{\partial p}dp.
\end{array}
\label{8}
\end{equation} 
Using Eqs.~(2), (8), and the relationship between the complex conductivity and
complex susceptibility $\chi _{ij}=\frac{i\sigma _{ij}}{\omega }$, we arrive
at%
\begin{equation}
m_{z}^{\left( 0\right) }=-\frac{g\pi e^{3}}{2c\omega ^{2}}%
\int_{0}^{\infty }\left( \frac{\partial W}{\partial p}\right) ^{2}\frac{%
\partial f_{F}}{\partial p}dp\times \mathrm{Re}\left( i\tilde{E}_{y}\tilde{E}%
_{x}^{\ast }\right) ,
\end{equation}
where the superscript $(0)$ indicates the transparent medium approximation
used to derive the Pitaevskii equation Eq.~(1).

Since the effect is strongest when the electrons are strongly
degenerate, we consider a zero-temperature 2D Fermi distribution as an
unperturbed electron distribution:
\begin{equation}
f_{F}\left( p\right) =\frac{1}{\left( 2\pi \hbar \right) ^{2}}\Theta \left(
p_{F}-p\right) ,
\end{equation}
where $\Theta \left( x\right) $ is the Heaviside step function. In this case
the integrals are easily calculated to give
\begin{equation}
\begin{array}{l}
\displaystyle \sigma _{xx} =\sigma _{yy}=\sigma =i\frac{ge^{2}p_{F}}{4\pi \hbar
^{2}\omega }\left( \frac{\partial W}{\partial p}\right) _{p=p_{F}},  
\\
\displaystyle \sigma _{xy} =-\sigma _{yx}=\frac{ge^{3}H_{z}}{4\pi c\hbar ^{2}\omega ^{2}}%
\left( \frac{\partial W}{\partial p}\right) _{p=p_{F}}^{2}. 
\end{array}
\end{equation}
In particular, for graphene with linear dispersion $(g=4,$ $\frac{\partial W%
}{\partial p}=v_{F})$ the last of Eqs.~(11) yields 
\begin{equation}
\label{12}
\sigma _{xy}^{\left( intra\right) }=\frac{e^{3}v_{F}^{2}H_{z}}{\pi c\hbar
^{2}\omega ^{2}}
\end{equation}%
Here we added the label (intra) to emphasize the fact that the
quasiclassical calculation gives only the intraband conductivity. For the
magnetic moment we obtain
\begin{equation}
m_{z}^{\left( 0\right) }=\frac{ge^{3}}{8\pi c\hbar ^{2}\omega ^{3}}\left(
\frac{\partial W}{\partial p}\right) _{p=p_{F}}^{2}\times \mathrm{Re}\left( i%
\tilde{E}_{y}\tilde{E}_{x}^{\ast }\right) .
\end{equation}%
It follows from Eq.~(13) that if the electron dispersion is quadratic, the
magnetization is proportional to the surface electron density $n_{F}=\frac{%
gp_{F}^{2}}{4\pi \hbar ^{2}}$ and inversely proportional to the square of
their effective mass. For a linear dispersion near the Dirac point as in
Eq.~(4) and degenerate electron distribution of Eq.~(10) the magnetization
does not depend on the Fermi momentum $p_{F}$, i.e. it does not depend on
the carrier density. One can write the result in the same form for both
cases by introducing an effective mass for electrons at the Fermi level in
graphene: $m_{\rm eff}$ $=\frac{p_{F}}{v_{F}}$. One has to keep in mind that the
limit of small $p_F \rightarrow 0$ is not allowed as it would violate not only the criterion of
negligible contribution from interband transitions but also the
applicability of the method of small perturbations that we used when solving
the kinetic equation. The latter condition has the form $p_{F}\gg \frac{%
eE_{0}}{\omega }$ , where $E_{0}=\left\vert \mathbf{\tilde{E}}\right\vert $,
as follows from the solution for the strong-field nonlinear problem solved in Appendix B.

\section{Quantum theory of the IFE in graphene}

The magnetic moment generated as a result of IFE is determined by the
magnetic field dependence of the off-diagonal element of the conductivity
tensor. To find this dependence within full quantum theory we use the
Kubo-Greenwood formula \cite{mahan2000}:%
\begin{equation}
\sigma _{xy}=-\sigma _{yx}=i\hbar g\sum_{\alpha \beta }\left( \frac{%
f_{\alpha }-f_{\beta }}{E_{\beta }-E_{\alpha }}\right) \frac{\left\langle
\alpha \right\vert \hat{\jmath}_{x}\left\vert \beta \right\rangle
\left\langle \beta \right\vert \hat{\jmath}_{y}\left\vert \alpha
\right\rangle }{\hbar (\omega +\frac{i}{\tau })-(E_{\beta }-E_{\alpha })},
\end{equation}%
where $\left\vert \alpha \right\rangle $ are basis 2D surface states
normalized by unit area $L_{x}\times $ $L_{y}=1$, $E_{\alpha }$ and $%
f_{\alpha }$ are the energy and population of state $\left\vert \alpha
\right\rangle $, $\hat{\jmath}_{x,y}=-ev_{F}\hat{\sigma}_{x,y}$ are
Cartesian components of the current density operator \cite{katsnelson2012}, $%
g=4$ is the degeneracy factor, $\tau $ is the relaxation time.

To determine the distribution function of carriers in a magnetic field
oriented along z-axis, we extend the momentum operator in the Hamiltonian
Eq.~(3) in a standard way \cite{landau1977}: $\mathbf{\hat{p}\Longrightarrow
\hat{p}}-\mathbf{x}_{0}\frac{eH_{z}}{c}y$ . The resulting electron
eigenstates are \cite{mcclure1956}

\begin{equation}
\left\vert \alpha \right\rangle =\left\vert n,k\right\rangle =\frac{C_{n}}{%
\sqrt{L_{y}}}e^{-ik_{y}y}\left(
\begin{array}{c}
\mathrm{sgn}\left( n\right) i^{\left\vert n\right\vert -1}\phi _{\left\vert
n\right\vert -1} \\
i^{\left\vert n\right\vert }\phi _{\left\vert n\right\vert }%
\end{array}%
\right)
\end{equation}
\begin{equation}
\phi _{\left\vert n\right\vert }=\frac{H_{\left\vert n\right\vert }\left(
\frac{x-kl_{c}^{2}}{l_{c}}\right) }{\sqrt{2^{\left\vert n\right\vert
}\left\vert n\right\vert !\sqrt{\pi }l_{c}}}\exp \left[ -\frac{1}{2}\left(
\frac{x-kl_{c}^{2}}{l_{c}}\right) ^{2}\right] ,
\end{equation}%
where $H_{n}\left( \xi \right) $ is the Hermite polynomial, $l_{c}=\sqrt{%
\frac{\hbar c}{eH_{z}}}$ is the magnetic length, $n=0,\pm 1,\pm 2,...$are
principal numbers of the Landau levels, $C_{0}=1$, $C_{n\neq 0}=\frac{1}{%
\sqrt{2}}$. The eigenenergy $E_{\alpha }$ depends only on the Landau level
number: $E_{\alpha }=E_{n}=\mathrm{sgn}\left( n\right) \hbar \omega _{c}$$%
\sqrt{\left\vert n\right\vert }$, where $\omega _{c}=\frac{\sqrt{2}v_{F}}{%
l_{c}}$ is the cyclotron frequency.

Introducing the notations $\left\vert \alpha \right\rangle =\left\vert
n,k\right\rangle $ and $\left\vert \beta \right\rangle =\left\vert
m,k^{\prime }\right\rangle $ and using Eqs.~(15) and (16) we obtain the
matrix elements of the components of the current density operator:
\begin{equation}
\left\langle \alpha \right\vert \hat{\jmath}_{x,y}\left\vert \beta
\right\rangle =-ev_{F}\left\langle \alpha \right\vert \hat{\sigma}%
_{x,y}\left\vert \beta \right\rangle =\left( j_{x,y}\right) _{nm}\delta
_{kk^{\prime }},
\end{equation}%
where%
\begin{equation}
(j_{x})_{nm}=-ev_{F}i^{\left\vert m\right\vert -\left\vert n\right\vert
+1}C_{n}C_{m}\left[ \mathrm{sgn}\left( n\right) \delta \left( \left\vert
n\right\vert -\left\vert m\right\vert -1\right) -\mathrm{sgn}\left( m\right)
\delta \left( \left\vert n\right\vert -\left\vert m\right\vert +1\right) %
\right] ,
\end{equation}%
\begin{equation}
(j_{y})_{nm}=-ev_{F}i^{\left\vert m\right\vert -\left\vert n\right\vert
}C_{n}C_{m}\left[ \mathrm{sgn}\left( m\right) \delta \left( \left\vert
n\right\vert -\left\vert m\right\vert +1\right) +\mathrm{sgn}\left( n\right)
\delta \left( \left\vert n\right\vert -\left\vert m\right\vert -1\right) %
\right].
\end{equation}%
The $\delta $-functions in Eqs.~(18), (19) determine the selection rules.

Performing the summation over $k$ in Eq.~(14) (see \cite{landau1977}) and
using Eqs.~(18),(19), we arrive at the expression which contains the
summation over the Landau level numbers:
\begin{equation}
\label{20}
\sigma _{xy}=-\frac{2\hbar }{\pi l_{c}^{2}}e^{2}v_{F}^{2}\sum_{mn}\left(
C_{n}C_{m}\right) ^{2}\frac{f_{n}-f_{m}}{E_{m}-E_{n}}\frac{\delta \left(
\left\vert n\right\vert -\left\vert m\right\vert -1\right) -\delta \left(
\left\vert n\right\vert -\left\vert m\right\vert +1\right) }{\hbar (\omega +%
\frac{i}{\tau })+(E_{n}-E_{m})}
\end{equation}%
where $1\geq f_{n}\geq 0$; the degeneracy of a given Landau level per unit
area is $\frac{2\hbar }{\pi l_{c}^{2}}$ including both spin and valley
degeneracy.

In the case of a complete electron-hole symmetry, i.e.$f_{0}=\frac{1}{2},\
f_{n>0}=0,$ $f_{n<0}=1,$ from Eq.~(20) we obtain $\sigma _{xy}\equiv 0$ for
any $H_{z}$ (see also \cite{long2018}). Now consider an n-doped system. Let
the number $n_{F}$ correspond to the highest occupied Landau level just
below the Fermi energy, i.e. $W_{F}$ $\geq \hbar \omega _{c}$$\sqrt{n_{F}}$.
Since we need the limit of small magnetic fields, we assume that $W_{F}$ $%
\gg \hbar \omega _{c}$, which can be written as
\begin{equation}
p_{F}l_{c}\gg \hbar. 
\end{equation}%
This means that $n_{F}\gg 1$.

\subsection{The contribution of intraband transitions}

In this case we put $n,m>0$ in Eq.~(\ref{20}). Consider a narrow vicinity of the
Fermi energy where $\left\vert n-n_{F}\right\vert \ll n_{F}$ and $\left\vert
E_{n_{F}}-W_{F}\right\vert \ll W_{F}$ . In the limit of large $n$ the
distance between neighboring Landau levels is
\begin{equation}
\Delta E=E_{n+1}-E_{n}=\hbar \omega _{c}\left( \sqrt{n+1}-\sqrt{n}\right)
\approx \frac{1}{2}\frac{\hbar \omega _{c}}{\sqrt{n_{F}}},
\end{equation}%
or%
\begin{equation}
\Delta E=\frac{\hbar ^{2}v_{F}^{2}}{l_{c}^{2}W_{F}}
\end{equation}%
Note that introducing the effective mass $m_{\rm eff}=\frac{p_{F}}{v_{F}}$ we
obtain a standard relation $\Delta E=\frac{\hbar eH_{z}}{cm_{\rm eff}}$ .

\qquad Taking into account that $f_{n+1}-f_{n}\neq 0$ only in the near
vicinity of the Fermi energy, from Eq.~(\ref{20}) we can get 
\begin{equation}
\sigma _{xy}^{\left( intra\right) }=-\frac{\hbar }{2\pi l_{c}^{2}} 
e^{2}v_{F}^{2}\frac{1}{\Delta E}\left[ \frac{1}{\hbar (\omega +\frac{i}{\tau
})-\Delta E}-\frac{1}{\hbar (\omega +\frac{i}{\tau })+\Delta E}\right]
\sum_{n>0}\left( f_{n+1}-f_{n}\right) ,
\end{equation} 
where $\sum_{n>0}\left( f_{n+1}-f_{n}\right) \Longrightarrow
\int_{0}^{\infty }df=-1$. The result is%
\begin{equation}
\label{intra}
\sigma _{xy}^{\left( intra\right) }=\frac{1}{\pi \hbar ^{2}c}\frac{%
e^{3}v_{F}^{2}H_{z}}{(\omega +\frac{i}{\tau })^{2}-\left( \frac{eH_{z}v_{F}}{%
cp_{F}}\right) ^{2}}.
\end{equation}%
The last expression coincides with the semiclassical result derived from the
kinetic equation Eq.~(6) for $\hat{Q}\left( f\right) =\frac{f_{F}-f}{\tau }$
. In particular, when $\tau \rightarrow \infty $ and $H_{z}\rightarrow 0$ we
obtain Eq.~(\ref{12}).

\subsection{ The contribution of interband transitions}

 In this case the numbers $n$ and $m$ in Eq.~(\ref{20}) have different
signs. Taking this into account, we can write the sum in Eq.~(\ref{20}) as%
\begin{eqnarray}
\sigma _{xy}^{\left( inter\right) } &=&\frac{-\hbar e^{2}v_{F}^{2}}{2\pi
l_{c}^{2}}[\sum_{n<0,m>0}\frac{f_{n}-f_{m}}{E_{m}+\left\vert
E_{n}\right\vert }\frac{\delta \left( n+m+1\right) -\delta \left(
n+m-1\right) }{\hbar (\omega +\frac{i}{\tau })-(\left\vert E_{n}\right\vert
+E_{m})}  \notag \\
&&-\sum_{n>0,m<0}\frac{f_{n}-f_{m}}{\left\vert E_{m}\right\vert +E_{n}}\frac{%
\delta \left( n+m-1\right) -\delta \left( n+m+1\right) }{\hbar (\omega +%
\frac{i}{\tau })+(E_{n}+\left\vert E_{m}\right\vert )}].
\label{26}
\end{eqnarray}%
Since in an n-doped degenerate system  $f_{n>n_{F}}=0$ , $%
f_{n\leqslant n_{F}}=1$, Eq.~(\ref{26}) yields%
\begin{eqnarray} &&
\sigma _{xy}^{\left( inter\right) } =-\frac{\hbar e^{2}v_{F}^{2}}{2\pi
\hbar ^{2}l_{c}^{2}}\times \notag \\ &&
\left(  \sum_{-\left( n_{F}+2\right) }^{-\infty }\frac{1}{\frac{%
E_{-n-1}+\left\vert E_{n}\right\vert }{\hbar }\left[ (\omega +\frac{i}{\tau }%
)-\frac{E_{-n-1}+\left\vert E_{n}\right\vert }{\hbar }\right] }%
-\sum_{-n_{F}}^{-\infty }\frac{1}{\frac{E_{-n+1}+\left\vert E_{n}\right\vert
}{\hbar }\left[ (\omega +\frac{i}{\tau })-\frac{E_{-n+1}+\left\vert
E_{n}\right\vert }{\hbar }\right] } \right.  \notag \\ && \left.
-\sum_{n_{F}+1}^{\infty }\frac{1}{\frac{\left\vert E_{-n+1}\right\vert
+E_{n}}{\hbar }\left[ (\omega +\frac{i}{\tau })+\frac{\left\vert
E_{-n+1}\right\vert +E_{n}}{\hbar }\right] }+\sum_{n_{F}+1}^{\infty }\frac{1%
}{\frac{\left\vert E_{-n-1}\right\vert +E_{n}}{\hbar }\left[ (\omega +\frac{i%
}{\tau })+\frac{\left\vert E_{-n-1}\right\vert +E_{n}}{\hbar }\right] }\right)
\label{27}
\end{eqnarray}%
Since the energy spectrum is symmetric, $\left\vert E_{-\left\vert
n\right\vert }\right\vert =E_{\left\vert n\right\vert }$, we can regroup
the terms on the rhs of Eq.~(\ref{27}) as%
\begin{equation*}
\begin{array}{l}
(...) =-\frac{2}{(\omega +\frac{i}{\tau })^{2}-\left( \frac{
E_{n_{F}+1}+\left\vert E_{-n_{F}}\right\vert }{\hbar }\right) ^{2}}
-\sum_{n_{F}+2}^{\infty }\frac{2}{(\omega +\frac{i}{\tau })^{2}-\left( \frac{
\left\vert E_{-n+1}\right\vert +E_{n}}{\hbar }\right) ^{2}} \\ 
+\sum_{n_{F}+1}^{\infty }\frac{2}{(\omega +\frac{i}{\tau })^{2}-\left( \frac{
\left\vert E_{-n-1}\right\vert +E_{n}}{\hbar }\right) ^{2}}
\end{array}
\end{equation*}%
It is easy to see that the sums on the rhs of the last equation cancel each
other, leaving only the first term which is the contribution of the
transition $-n_{F}\Longrightarrow n_{F}+1$ (see Fig.~2). Taking into account
that $\frac{E_{n_{F}+1}+\left\vert E_{-n_{F}}\right\vert }{\hbar }\approx
\frac{2W_{F}}{\hbar }$ when the inequality Eq.~(21) is satisfied, we obtain%
\begin{equation}
\label{inter}
\sigma _{xy}^{\left( inter\right) }=\frac{1}{\pi \hbar ^{2}c}\frac{%
e^{3}v_{F}^{2}H_{z}}{(\omega +\frac{i}{\tau })^{2}-\left( \frac{2W_{F}}{%
\hbar }\right) ^{2}}
\end{equation}

In the absence of dissipation the magnitude of the magnetic moment is
determined by Eq.~(2), which gives
\begin{equation}
m_{z}^{\left( 0\right) }=\frac{1}{2\omega }\left[ \frac{\partial \left(
\sigma _{xy}^{\left( intra\right) }+\sigma _{xy}^{\left( inter\right)
}\right) }{\partial H_{z}}\right] _{\tau \rightarrow \infty
,H_{z}\rightarrow 0}\mathrm{Re}\left( i\tilde{E}_{y}\tilde{E}_{x}^{\ast
}\right) .
\end{equation}

Using Eqs.~(\ref{intra}) and (\ref{inter}) we finally arrive at 
\begin{equation}
m_{z}^{\left( 0\right) }=  \frac{e^{3}v_{F}^{2}}{2\pi c\hbar ^{2}\omega ^{3}}  \frac{\left( \frac{2W_{F}}{\hbar }\right)
^{2}-2\omega ^{2}}{\left( \frac{2W_{F}}{\hbar }\right) ^{2}-\omega ^{2}}%
\mathrm{Re}\left( i\tilde{E}_{y}\tilde{E}_{x}^{\ast }\right). 
\label{m}
\end{equation}

\begin{figure}[htb]
\begin{center}
\includegraphics[scale=0.4]{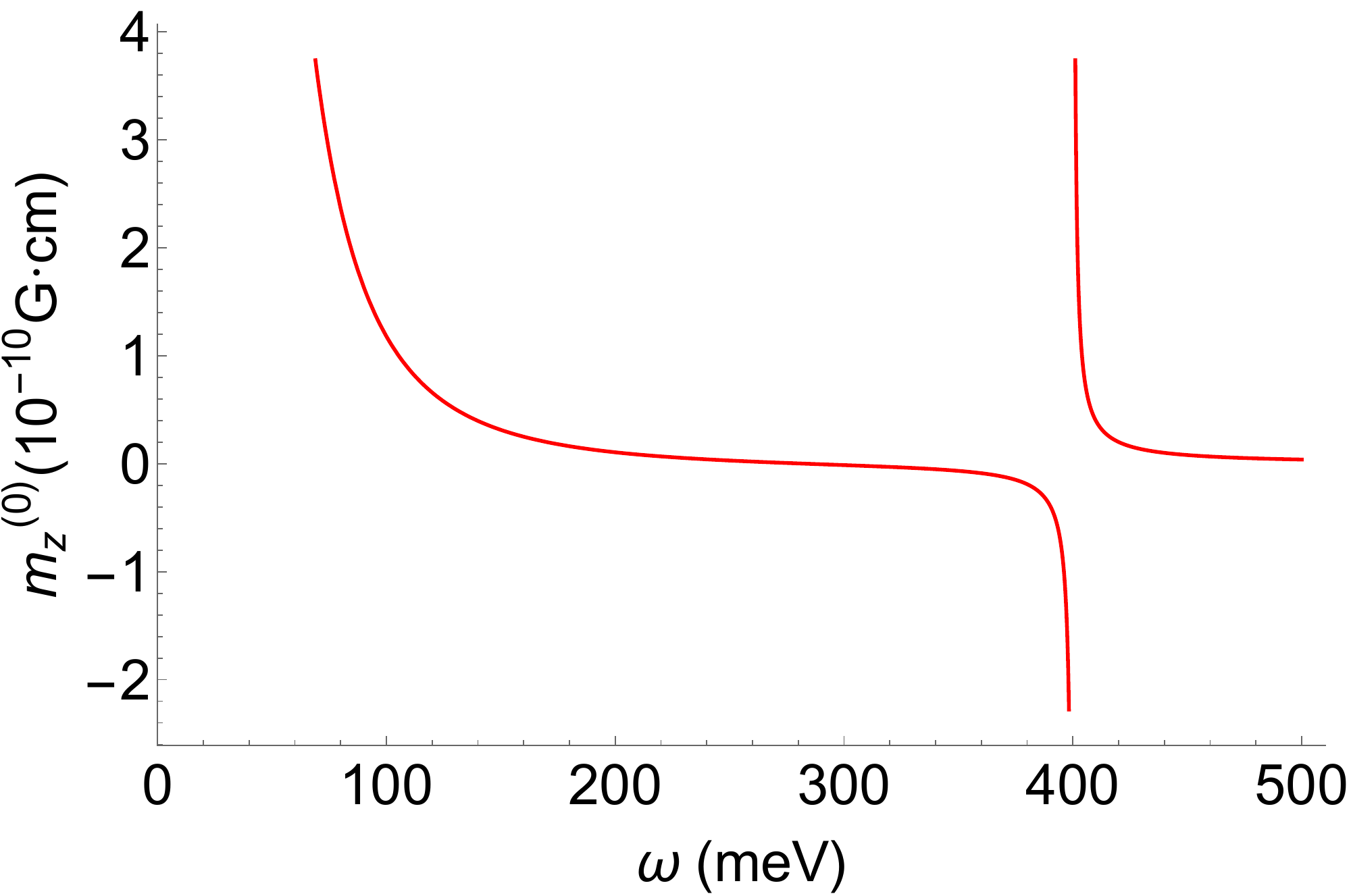}
\caption{Frequency dependence of the  magnetization in Eq.~(\ref{m}) induced by a circularly polarized optical field of intensity $10$ kW/cm$^2$. The Fermi energy $W_F = 0.2$ eV.  }
\label{fig3}
\end{center}
\end{figure}

The frequency dependence of the magnetization is shown in Fig.~3. The incident light intensity was assumed to be 10 kW/cm$^2$, which is much less than the saturation intensity, so that the contribution of photoexcited carriers can be neglected. The magnitude of magnetization increases with decreasing frequency as $1/\omega^3$ when $\hbar \omega \ll W_F$ and the effect is dominated by intraband transitions. The magnetization changes sign twice: at $\hbar \omega = \sqrt{2} W_F$ and $\hbar \omega = 2 W_F$. There is also a resonance at the interband transition edge $\hbar \omega = 2 W_F$ where the magnitude of magnetization diverges. The divergence is an artifact of the dissipationless approximation which was used to relate magnetization to the off-diagonal susceptibility elements in Eq.~(2). Obviously, relaxation processes cannot be neglected near resonance. Therefore the validity of Eq.~(\ref{m}) in the resonance region is limited by $|\omega - 2 W_F/\hbar| > \tau_{inter}^{-1}$, where $\tau_{inter}$ is the interband relaxation time.  It is interesting that taking relaxation processes into account in the calculation of magnetization is not equivalent to using the complex susceptibility in Eq.~(2) and taking the real part of the resulting expression. We will illustrate it in the next section within semiclassical derivation.

As is clear from Fig.~3 and Eqs.~(\ref{intra}), (\ref{inter}), and (\ref{m}), when Eq.~(5) is satisfied the interband transitions
give only a small contribution to the IFE. In the analysis of the IFE in
dissipative systems below, we will therefore neglect interband transitions.

\section{IFE in a dissipative system}

Here we calculate the photogenerated magnetic moment per unit area without
any assumptions of a dissipationless system. First we introduce surface
polarization $\mathbf{P}$ and relate it with the surface current $\mathbf{j}$
in a standard way $\mathbf{\dot{P}=j}$. Next, we represent polarization as $%
\mathbf{P=-}en_{F}\mathbf{R}$, where the vector $\mathbf{R}$ has a
meaning of an average displacement of carriers and $n_{F}$ is the surface
density of a degenerate 2D electron gas. The magnetic moment per unit area
is $\mathbf{m=-}n_{F}\times \frac{e}{2c}\left\langle \mathbf{R\times
\dot{R}} \right\rangle $, where the angular brackets mean averaging
over the optical period $\frac{2\pi }{\omega }$. This expression is
convenient to write as%
\begin{equation}
\label{31}
\mathbf{m=z}_{0}m_{z}=- \frac{1}{2cen_{F}}\left\langle 
\mathbf{P\times j} \right\rangle .
\end{equation}%
Substituting 
\begin{equation}
\label{32}
\mathbf{j=}\mathrm{Re}\left( \sigma \left( \omega \right) \mathbf{\tilde{E}}%
e^{-i\omega t}\right) ,\ \ \ \mathbf{P=}\mathrm{Re}\left( \frac{i}{\omega }%
\sigma \left( \omega \right) \mathbf{\tilde{E}}e^{-i\omega t}\right)
\end{equation}%
into Eq.~(\ref{31}), we obtain%
\begin{equation}
\label{mz}
m_{z}=\frac{\left\vert \sigma \left( \omega \right) \right\vert ^{2}}{%
2ce\omega n_{F}}\mathrm{Re}\left( i\tilde{E}_{y}\tilde{E}_{x}^{\ast }\right),
\end{equation}
where $\sigma = \sigma_{xx} = \sigma_{yy}$; see Eq.~(\ref{8}). 
For a classical plasma
Eq.~(\ref{mz}) was derived in \cite{mdtokman 1984}.

To connect with the dissipationless limit in Eq.~(2) we note that the
elements of the conductivity tensor given by Eqs.~(11) in a dissipationless
system for any electron dispersion are related as 
\begin{equation}
\label{34}
\frac{1}{ecn_{F}}\frac{i}{\omega }\left\vert \sigma \right\vert ^{2}=\left(
\frac{i}{\omega }\frac{\partial \sigma _{xy}}{\partial H_{z}}\right)
_{H_{z}\rightarrow 0}=\left( \frac{\partial \chi _{xy}}{\partial H_{z}}%
\right) _{H_{z}\rightarrow 0}
\end{equation}%
Substituting this into Eq.~(\ref{mz}), we obtain the expression for magnetization
which coincides with the phenomenological formula of Eq.~(2).

Therefore, an approach based on Eqs.~(\ref{31}) and (\ref{32}) which uses the conductivity $
\sigma \left( \omega \right) $ calculated within a suitable microscopic
model, leads to a correct result. Note that this approach is not based on
dissipationless approximation. An advantage of an approach based on Eq.~(\ref{31})
is that there is no need to calculate the dielectric susceptibility tensor
in the limit of a linear dependence on the external magnetic field $\mathbf{H%
}$. It is enough to calculate linear conductivity without an external
magnetic field. In order to include dissipation, we use Eq.~(6), assuming $%
\mathbf{H=0}$ from the very beginning and adopting the simplest
approximation for the relaxation operator: $\hat{Q}\left( f\right) =\frac{%
f_{F}-f}{\tau }$, where $\tau $ is the relaxation time. This is equivalent
to the substitution $\omega \rightarrow \omega +\frac{i}{\tau }$ in the
dissipationless formula. Then Eq.~(\ref{mz}) gives%
\begin{equation}
\label{diss} 
m_{z}=m_{z}^{\left( 0\right) }\frac{\omega ^{2}}{\omega ^{2}+\tau ^{-2}}
\end{equation}%
where $m_{z}^{\left( 0\right) }$ is the magnetization of a dissipationless
system, see Eq.~(13).  One can see that Eq.~(\ref{diss}) is not equivalent to using the complex susceptibility in Eq.~(2) and taking the real part of the resulting expression. 

At low frequencies, the finite size of a sample starts affecting the
result; see Appendix A. 
The expression for the magnetic moment which is valid beyond the linearized
theory is derived in Appendix B.

\section{ The magnetization current and finite-size effects}

The magnetization current density generated in a 2D system as a result of
IFE is given by $\mathbf{j=}c\left( \mathbf{x}_{0}\frac{\partial m_{z}}{%
\partial y}-\mathbf{y}_{0}\frac{\partial m_{z}}{\partial x}\right) $. This
equation yields a simple expression for the photocurrent around the boundary
of a light beam or along the edge of an illuminated sample:%
\begin{equation}
\label{bound}
\mathbf{I=}c\left[ \mathbf{n}_{0}\times \mathbf{z}_{0}\right] m_{z},
\end{equation}
where $\mathbf{n}_{0}$ s a unit vector in the monolayer plane which is
directed outside from the illuminated area perpendicularly to the boundary, 
see Fig.~4.


\begin{figure}[htb]
\begin{center}
\includegraphics[scale=0.4]{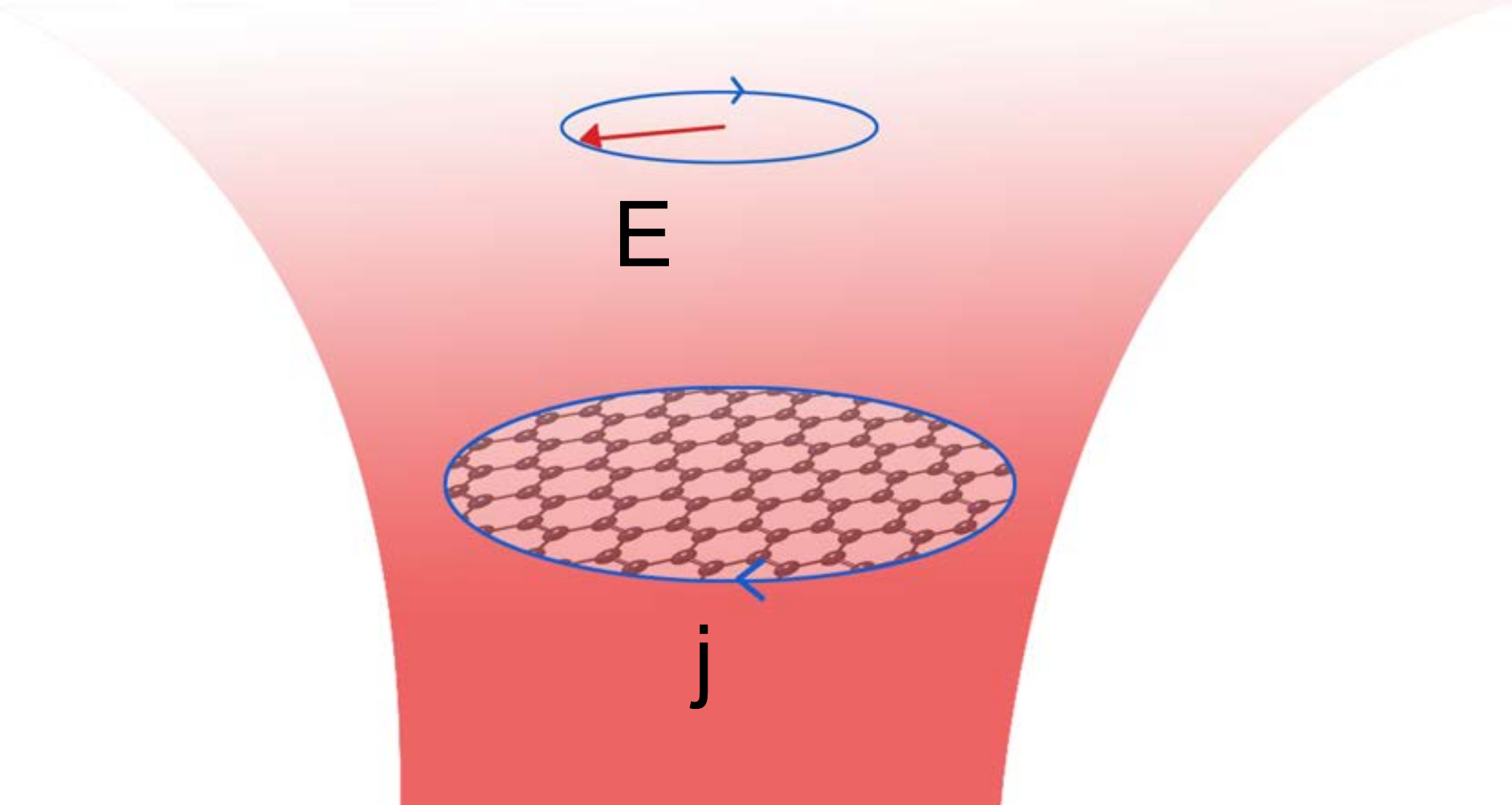}
\caption{A sketch of an edge photocurrent in a finite-size sample generated by an incident circularly polarized beam. }
\label{fig4}
\end{center}
\end{figure}


In a dissipative system a simple expression Eq.~(\ref{bound}) may be used 
with certain reservations. For example, the magnetization current far from
the sample edges can be affected by the viscosity of an electron fluid \cite%
{levitov2016} whereas edge photocurrent can be affected by interaction of
carriers with a sample boundary. (These effects can be responsible for various
ways of detecting a constant current along the edge that are not related to
IFE.)  In fact, Eq.~(\ref{bound}) corresponds to a mirror reflection of carriers from
the boundary. Indeed, consider the edge $x=0$ of a graphene sample, assuming
that graphene extends to $x>0$. The field component $E_{x}=\mathrm{Re}\left(
\tilde{E}_{x}e^{-i\omega t}\right) $ excites oscillations of carrier density
in a transition layer near the edge: $\delta n\left( x\right) =\mathrm{Re}%
\left( \delta \tilde{n}\left( x\right) e^{-i\omega t}\right) $. Oscillations
of an uncompensated charge $\delta \rho =-e\delta n$ should obey the
continuity equation, which gives%
\begin{equation}
\label{35}
i\omega e\int_{0}^{\infty }\delta \tilde{n}\left( x\right) dx=\sigma \left(
\omega \right) \tilde{E}_{x},
\end{equation}%
where the conductivity $\sigma \left( \omega \right) $ corresponds to the
region where there is no uncompensated charge. Although the integration here should be formally extended to $x\rightarrow \infty $, in practice
it is localized within a certain transition layer much smaller than the sample
dimensions.

The field component $E_{y}=\mathrm{Re}\left( \tilde{E}_{y}e^{-i\omega
t}\right)$ gives rise to the oscillations of carrier velocity along the
edge. We can prove that for the elastic reflection of electrons from the
boundary the average (hydrodynamic) velocity of electrons along the boundary
(along y) is conserved up to cubic terms with respect to the field
amplitude. Indeed, let us write the particle momentum as $\mathbf{p}=\mathbf{%
P}+\mathbf{\tilde{p}}\left( t\right) $ , where $\mathbf{P}$ is its value
averaged over time and $\mathbf{\tilde{p}}=\frac{e}{\omega }\mathrm{Re}%
\left( i^{-1}\mathbf{\tilde{E}}e^{-i\omega t}\right) $ is an oscillating
component. The velocity $\mathbf{v=}v_{F}$ $\frac{\mathbf{p}}{\left\vert
\mathbf{p}\right\vert }$ in the linear approximation with respect to the
field $\mathbf{E}$ is given by $\mathbf{\tilde{v}}$ $\approx v_{F}$ $\left(
\frac{\mathbf{\tilde{p}}}{\left\vert \mathbf{P}\right\vert }-\frac{\mathbf{P}%
\left( \mathbf{P}\cdot \mathbf{\tilde{p}}\right) }{\left\vert \mathbf{P}%
\right\vert ^{3}}\right) $, which gives%
\begin{equation}
\label{37}
\tilde{v}_{y}=v_{F}\left( \tilde{p}_{y}\frac{P_{x}^{2}}{\left(
P_{x}^{2}+P_{y}^{2}\right) ^{\frac{3}{2}}}-\tilde{p}_{x}\frac{P_{y}P_{x}}{%
\left( P_{x}^{2}+P_{y}^{2}\right) ^{\frac{3}{2}}}\right) .
\end{equation}%
If the particle distribution is symmetric with respect to $P_{y}$, the
ensemble-averaged velocity obtained from Eq.~(\ref{37}) is%
\begin{equation}
\label{38}
\left\langle v_{y}\right\rangle =V_{y}=v_{F}\tilde{p}_{y}\frac{\left\langle
P_{x}^{2}\right\rangle }{\left( P_{x}^{2}+P_{y}^{2}\right) ^{\frac{3}{2}}}.
\end{equation}%
For elastic reflection the momentum components $P_{y}$ and $\tilde{p}_{y}$ 
are conserved separately whereas the magnitude of $P_{x}^{2}$ changes upon
reflection. If $P_{x1}$ and $P_{x2}$ are the values before and after the
reflection, then $P_{x2}=-\left[ P_{x1}+2\tilde{p}_{x}\left( t^{\ast
}\right) \right] $ , where $t^{\ast }$ is the moment when the particle hits
the edge. If the phases $\omega t^{\ast }$ are uniformly distributed, this
effect contributes with the terms of the order of $\left\vert \tilde{E}%
_{x}\right\vert ^{2}$, which leads to corrections cubic with respect to the
field amplitude in Eq.~(\ref{38}). Neglecting these terms and also any effects of
viscosity in the transition layer we obtain $V_{y}=\mathrm{Re}\left( \tilde{V%
}_{y}e^{-i\omega t}\right) $, where $\tilde{V}_{y}=$ const. The result is 
\begin{equation}
\label{vy}
\tilde{V}_{y}=\tilde{V}_{y}\left( \infty \right) =\frac{\sigma \left( \omega
\right) E_{y}}{-en_{F}}.
\end{equation}%
Now we can calculate the constant (time-averaged) nonlinear edge
photocurrent as 
\begin{equation}
I_{y}=-\frac{e}{2}\mathrm{Re}\int_{0}^{\infty }\tilde{V}_{y}\delta \tilde{n}%
^{\ast }\left( x\right) dx.
\end{equation}%
Substituting here Eqs.~(\ref{35}),(\ref{vy}) yields 
\begin{equation}
\label{42}
I_{y}=\frac{1}{2en_{F}}\mathrm{Re}\frac{i}{\omega }\left\vert \sigma \left(
\omega \right) \right\vert ^{2}\tilde{E}_{y}\tilde{E}_{x}^{\ast }.
\end{equation}%
This result is exactly the same as the substitution of Eq.~(\ref{mz}) into Eq.~(\ref{bound}).

In the case of a very strong dissipation, when carriers are thermalized near
the edge, one calculates the edge current using the approach described in
\cite{glazov2014}. This method relates the perturbation of carrier density
with the perturbation of the chemical potential in the Fermi distribution.
Applying this approach to a 2D system with linear electron dispersion gives
the result which differs from Eq.~(\ref{42}) by a factor of $\frac{1}{2}$, whereas
in a 3D with linear dispersion system the difference is a factor of $\frac{2%
}{3}$. In materials with a constant effective mass the result is the same as
Eq.~(\ref{42}). Note that in graphene and in typical semiconductors the
thermalization time for carriers in a given band is longer than their
scattering time by at least one order of magnitude; see e.g. \cite%
{oladyshkin2017} and references therein. For a model with diffuse scattering
at the boundary \cite{karch2011}, the expression in Eq.~(\ref{42}) gives only an order
of magnitude estimate.

\section{IFE in Weyl semimetals}

We consider the simplest model of a Dirac or Weyl Type I semimetal
(hereafter WSM) valid only at low enough frequencies in the near vicinity of a Weyl point, which is
basically a 3D generalization of Eqs.~(3),(4), in which $\mathbf{\hat{p}}$ is
a 3D momentum operator, $\boldsymbol{\hat{\sigma}} =\mathbf{x}_{0}\hat{\sigma}%
_{x}+\mathbf{y}_{0}\hat{\sigma}_{y}+\mathbf{z}_{0}\hat{\sigma}_{z}$ is a 3D
vector of Pauli matrices, and
\begin{equation}
\label{43}
W(p)=sv_{F}\sqrt{p_{x}^{2}+p_{y}^{2}+p_{z}^{2}}.
\end{equation}%
Here the number of Weyl nodes only adds to the degeneracy of electron states and the optical anisotropy and gyrotropy effects related to the finite separation of Weyl nodes \cite{chen2019} are neglected. The volume conductivity can be derived from a single-band kinetic equation
if the radiation frequency $\omega $, Fermi energy $v_{F}p_{F}$ and the
distance $b$ between Weyl nodes in $k$-space are related by \cite{chen2019}:%
\begin{equation*}
\hbar \omega \ll v_{F}p_{F}\ll \hbar v_{F}b
\end{equation*}%
For an unperturbed Fermi distribution in the conduction band, 
\begin{equation}
\label{44}
f_{F}\left( p\right) =\frac{1}{\left( 2\pi \hbar \right) ^{3}}\Theta \left(
p_{F}-p\right) ,
\end{equation}
the conductivity has a Drude-like form \cite{chen2019}:%
\begin{equation}
\label{45}
\sigma =i\frac{e^{2}n_{F}}{\omega +\frac{i}{\tau }}\times \frac{v_{F}}{p_{F}},
\end{equation}%
where $n_{F}=$ $\frac{gp_{F}^{3}}{6\pi ^{2}\hbar ^{3}}$ is a volume density
of electrons corresponding to the Fermi distribution Eq.~(\ref{44}); the degeneracy
$g$ takes into account the contribution of all Weyl nodes, including those
with opposite chiralities.

First consider the collisionless limit. We can again use Eq.~(6), taking $%
\hat{Q}\left( f\right) =0$ and $\mathbf{E\perp H\parallel z}_{0}$. For a 3D
system the solution to Eq.~(6) can be sought as $f=\mathrm{Re}\left[ \delta
f\left( \theta ,\phi ,p\right) e^{-i\omega t}\right] +f_{F}\left( p\right) ,$
where $p_{x}=p\cos\theta \sin\phi ,p_{y}=p\sin\theta \sin\phi
,p_{z}= p \cos\phi ;\left\vert \delta f\right\vert \ll f_{F}.$ Linearizing
Eq.~(6) and taking into account electron dispersion Eq.~(\ref{43}) gives%
\begin{equation}
\label{46}
-i\omega \delta f+\frac{v_{F}}{p}\frac{eH_{z}}{c}\frac{\partial \delta f}{%
\partial \theta }-e\left( \tilde{E}_{x}\cos \theta +\tilde{E}_{y}\sin \theta
\right) \sin \phi \frac{\partial f_{F}}{\partial p}=0.
\end{equation}%
Eq.~(\ref{46}) has the following solution:%
\begin{equation}
\label{47}
\delta f=\frac{e}{\omega ^{2}-\left( \frac{v_{F}}{p}\frac{eH_{z}}{c}\right)
^{2}}\frac{\partial f_{F}}{\partial p}\sin \phi \left[ \tilde{E}_{x}\left(
i\omega \cos \theta -\frac{v_{F}}{p}\frac{eH_{z}}{c}\sin \theta \right) +%
\tilde{E}_{y}\left( i\omega \sin \theta +\frac{v_{F}}{p}\frac{eH_{z}}{c}\cos
\theta \right) \right] .
\end{equation}%
The corresponding current density is%
\begin{eqnarray}
j_{x} &=&-egv_{F}\mathrm{Re}\left( e^{-i\omega t}\int \sin \phi \cos \theta
\delta fd^{3}p\right) ,  \notag \\
j_{y} &=&-egv_{F}\mathrm{Re}\left( e^{-i\omega t}\int \sin \phi \sin \theta
\delta fd^{3}p\right).
\label{48}
\end{eqnarray}%
From Eqs.~(\ref{47}) and (\ref{48}) one can obtain the components of the conductivity tensor,
keeping only the terms linear with respect to the magnetic field:
\begin{eqnarray}
\sigma _{xx} &=&\sigma _{yy}=\sigma =\frac{4\pi ie^{2}gv_{F}}{3\omega }%
\int_{0}^{\infty }2f_{F}pdp,  \notag \\
\sigma _{xy} &=&-\sigma _{yx}=\frac{4\pi e^{3}gH_{z}v_{F}^{2}}{3\omega ^{2}c}%
\int_{0}^{\infty }f_{F}dp.
\label{49}
\end{eqnarray}
This gives the desired components of the dielectric permittivity tensor, $%
\varepsilon _{ij}=\delta _{ij}+4\pi \frac{i\sigma _{ij}}{\omega }$, and
finally the magnetic moment calculated using Eq.~(1):%
\begin{eqnarray}
m_{z}^{\left( 0\right) } &=& \frac{1}{8\pi }\mathrm{Re}\left[ \left(
\frac{\partial \varepsilon _{xy}^{\left( intra\right) }}{\partial H_{z}}%
\right) _{H_{z}\rightarrow 0}\tilde{E}_{y}\tilde{E}_{x}^{\ast }\right]
\notag \\
&=&\frac{2\pi e^{3}gH_{z}v_{F}^{2}}{3\omega ^{2}c}\int_{0}^{\infty
}f_{F}dp\times \mathrm{Re}\left( i\widetilde{E}_{y}\widetilde{E}_{x}^{\ast
}\right) ,
\end{eqnarray}
where the superscript $(0)$ is again to indicate an approximation of a
transparent medium.

For a degenerate electron distribution in the zero-temperature limit
Eq.~(\ref{44}) we have%
\begin{eqnarray}
\sigma _{xx} &=&\sigma _{yy}=\sigma =i\frac{e^{2}gp_{F}^{2}v_{F}}{6\hbar
^{3}\pi ^{2}\omega },  \notag \\
\sigma _{xy} &=&-\sigma _{yx}=\frac{e^{3}gH_{z}p_{F}v_{F}^{2}}{6\hbar
^{3}\pi ^{2}\omega ^{2}c},
\end{eqnarray}
and
\begin{equation}
\label{mzweyl}
m_z^{(0)} = \frac{e^3 gp_F v_F^2}{12 \hbar^3 \pi^2 \omega^3 c} {\rm Re}\left(i \tilde{E}_{y}\tilde{E}_{x}^{\ast } \right).
\end{equation}

As in the case of a 2D material, these components of the conductivity tensor
coincide with those obtained for particles with a constant mass $m_{\rm eff}$,
if we express them through a particle density $n_{F}$ and introduce the
effective mass as $m_{\rm eff}=\frac{p_{F}}{v_{F}}$.

It is also easy to find out that Eqs.~(\ref{49}) satisfy the equations similar to
those for 2D systems in Eq.~(\ref{34}):%
\begin{equation}
\frac{1}{ecn_{F}}\frac{i}{\omega }\left\vert \sigma \right\vert ^{2}=\left(
\frac{i}{\omega }\frac{\partial \sigma _{xy}}{\partial H_{z}}\right)
_{H_{z}\rightarrow 0}=\frac{1}{4\pi }\left( \frac{\partial \varepsilon _{xy}%
}{\partial H_{z}}\right) _{H_{z}\rightarrow 0}. 
\end{equation}

When scattering and dissipation are taken into account, one can repeat the
same derivation steps as above for a 2D system and arrive at the expression
for the photogenerated magnetic moment in the form of Eq.~(\ref{mz}), in which one
should substitute the volume conductivity Eq.~(\ref{45}) and volume carrier density
$n_{F}$.

\section{Discussion}

In order to compare the magnitude of the IFE in Dirac materials with that in conventional semiconductors, we note that for materials with conventional quadratic dispersion of carriers the induced magnetic moment per free carrier scales inversely proportional to their effective mass squared. As we already pointed out, the same dependence exists in both 2D and 3D Dirac materials if we denote $m_{\rm eff} = \frac{p_F}{v_F} = \frac{W_F}{v_F^2}$ as an effective mass. Assuming $v_F \approx c/300$, the ratio of the effective to free electron mass is $\frac{m_{\rm eff}}{m_0}  \simeq 2 \times 10^{-4} \frac{W_F}{1\; {\rm meV}}$. For example, when $W_F = 50$ meV, the effective mass is 0.01 $m_0$, which is one order of magnitude lower that in a typical semiconductor with a bandgap of the order of 1 eV. Therefore, at low frequencies $\hbar \omega \ll W_F$ the IFE in Dirac materials can be stronger than in conventional semiconductors by a couple of orders of magnitude. 

Let us estimate the magnetization obtained in the experiment \cite{karch2011}, where the excitation of edge photocurrent in graphene was investigated.
They used an NH$_{3}$ laser with 10 kW power and minimum frequency of 1.1
THz. For a 1 mm radius of a laser focus and Fermi energy of 0.2-0.3 eV the
condition $p_{F}\gg \frac{eE_{0}}{\omega }$ is satisfied. Using the
current dissipation time $\tau \sim 100$ fs (which corresponds to $\omega \tau \sim
1$), the magnetic moment of an illuminated spot is about $\sim
10^{-7}$ G cm$^{3}$, and the photoinduced average magnetic moment per
free carrier particle is of the order of 100 Bohr magnetons.  

If the optical pumping creates the magnetic moment of 100 Bohr magnetons per carrier, the magnetic moment per unit area of graphene scales as  $ 4 \pi m_z \sim 10^{-5} \left( \frac{W_F}{100\; {\rm meV}}\right)^2$ G cm. Similarly, the magnetic moment per unit volume in an illuminated volume of a Weyl semimetal sample scales roughly as 
$ 4 \pi m_z \sim 2.2 g \left( \frac{W_F}{100\; {\rm meV}}\right)^3$ G, where $g$ is degeneracy including the total number of Weyl nodes.

One possible application for the IFE is to provide all-optical modulation of the polarization of the probe light transmitted through (or reflected from) an area of the optical excitation. For example, a probe light passing along $z$-axis through the area of optically induced magnetization $m_z$ experiences {\it direct} Faraday effect. The magnitude of the polarization rotation $\chi$ can be calculated using textbook Faraday effect formulas in which an external magnetic field $B_z$ is replaced by $4 \pi m_z$, where $m_z$ is an optically induced magnetic moment per unit volume:
\begin{equation} 
\chi(L) = \int_0^L \alpha dz,
\end{equation}
where 
\begin{equation}
\label{alpha}
\alpha = \frac{\omega}{2c} (n_O - n_X)
\end{equation}
and $n_{O,X}$ 
 are refractive indices of normal EM modes, i.e.~ordinary and extraordinary modes. In the simplest case of a dielectric tensor with $\varepsilon_{xx} = \varepsilon_{yy}$ the normal modes are circularly polarized and  
\begin{equation}
\label{normal}
n_{O,X}^2 = \varepsilon_{xx} \pm |\varepsilon_{xy}|,
\end{equation}
where $\varepsilon_{xx} = \varepsilon_{yy} = 1 + 4 \pi i \sigma/\omega$. For small magnetic fields $\varepsilon_{xy} \propto B_z$, so Eqs.~(\ref{alpha}) and (\ref{normal}) give
\begin{equation}
\label{alpha2}
\alpha \approx \frac{\omega}{2c \sqrt{\varepsilon_{xx}}}|\varepsilon_{xy}| \approx  \frac{\omega}{2c \sqrt{\varepsilon_{xx}}} \left| \left( \frac{\partial \varepsilon_{xy}}{\partial B_z} \right)_{B_z \rightarrow 0} B_z   \right|. 
\end{equation}
Note that for the material with no intrinsic magnetic order and for linear dependence of the off-diagonal component of the dielectric tensor on the magnetic field, we can replace the magnetic field $H_z$ with the magnetic induction $B_z$ in all expressions in this paper. Then, taking into account that 
$$ m_z = \frac{1}{8 \pi} \left( \frac{\partial \varepsilon_{xy}}{\partial B_z} \right)_{B_z \rightarrow 0} |E|^2,$$
we obtain 
\begin{equation}
\label{alpha3}
\alpha = \frac{\omega}{4c} \frac{1}{\sqrt{1 + 4 \pi i \sigma/\omega}} \left( \frac{\partial \varepsilon_{xy}}{\partial B_z} \right)^2_{B_z \rightarrow 0} |E|^2,
\end{equation}
where 
$$ \sigma = i \frac{e^2 g v_F p_F^2}{6\hbar^3 \pi^2 \omega}, $$
$$ \left| \frac{\partial \varepsilon_{xy}}{\partial B_z} \right|_{B_z \rightarrow 0} = \frac{4 \pi}{\omega} \left| \frac{\partial \sigma_{xy}}{\partial B_z} \right|_{B_z \rightarrow 0} = \frac{2e^3 g v_F^2 p_F}{3\hbar^3 \pi \omega^3 c}. $$

For a specific example, consider an incident optical pump with the electric field of magnitude 10 kV/cm at frequency $\omega/2 \pi = 1$ THz. For the Fermi energy of 100 meV in a WSM sample the Faraday rotation parameter $\alpha \approx 6.6 g^{3/2}$ rad/cm, which is already interesting for applications.

\begin{acknowledgments} 
This work has been supported in part by the Air Force Office for Scientific Research
through Grant No.~FA9550-17-1-0341 and by NSF Award No.~1936276.
M.T. acknowledges the support from RFBR Grant No.~18-29-19091mk. 
I.O. acknowledges the support from Federal Research Center Institute of Applied Physics of the Russian Academy of Sciences (Project No. 0035-2019-004). I.T., I.S and V.P acknowledge the support by RFBR Grants No 18-02-00390, 19-31-51019 and the Russian State Contract No.~0035-2019-0021.
\end{acknowledgments}


\appendix

\section{Finite sample effects and the depolarization field}

Consider a sample shaped as a thin disk of radius $R$ in the $%
(x,y)$ plane and introduce polar coordinates $r$ and $\varphi $ on the disk. Consider a
circularly polarized optical field incident on a disk, with electric field
vector components
\begin{equation}
E_{x}=E_{0}\cos \left( \omega t\right) ,\ \ \ E_{y}=-E_{0}\sin \left( \omega
t\right),
\end{equation}
where $\omega >0$ corresponds to the clockwise rotation of the vector $%
\mathbf{E}$ and $\omega <0$ to the counterclockwise rotation. The rotating
field excites a rotating current in the disk:
\begin{equation}
\label{a2}
j_{x}=j_{0}\cos \left( \omega t+\phi \right) ,\ \ \ j_{y}=-j_{0}\sin \left(
\omega t+\phi \right),
\end{equation}
where the phase shift $\phi $ is determined by dissipative processes in the
sample. The current given by Eqs.~(A2) corresponds to the rotating electric
polarization:%
\begin{equation}
P_{x}=P_{0}\sin \left( \omega t+\phi \right) ,\ \ \ P_{y}=P_{0}\cos \left(
\omega t+\phi \right) ,
\end{equation}

where $P_{0}=\frac{j_{0}}{\omega }$, i.e. $\dot{P}_{x}=j_{x}$, $\dot{P}%
_{y}=j_{y}$.

The current excitation by a time-dependent external field in a finite sample
leads to an uncompensated time-dependent charge at a certain distance $l$
from the disk edge. The magnitude of the charge depends on the specific mechanism of
interaction of carriers with a boundary. Strictly speaking, both the current
and the electric polarization are described by Eqs.~(A2),(A3) only at a
certain distance $\rho \geq l$ from the disk edge. Since we don't want to
get into the details of the carrier-boundary interaction, we will assume that
the width of the boundary layer is much smaller than the disk radius: $l\ll
R $.

Let's denote an uncompensated charge per unit length along the disk edge as $%
\delta \rho \left( t,\varphi \right) $. It can be expressed as $\delta \rho
=P_{r}$ , where $P_{r}$ is the normal component of the polarization vector: $%
P_{r}=P_{x}\cos \varphi +P_{y}\sin \varphi $. The edge charge leads to
generation of the depolarization field $\mathbf{E}_{p}$ \cite{landau1984}. For a uniform
external field given by Eqs.~(A1), we can use the solution of a corresponding
electrostatic problem in \cite{landau1984}. If we approximate a thin disk with an
ellipsoid of rotation with semiminor axis $a\ll R$, we get%
\begin{equation*}
\mathbf{E}_{p}=-\frac{\pi ^{2}}{2R}\mathbf{P,}
\end{equation*}%
where $\mathbf{P}$ is a 2D density of the dipole moment. Taking into account
the effect of the depolarization field and Eqs.~(A1)-(A3), we obtain%
\begin{equation*}
\sigma \left[ E_{0}-i\frac{\pi ^{2}}{2R\omega }j_{0}e^{-i\phi }\right]
=j_{0}e^{-i\phi },~\ j_{0}e^{-i\phi }=E_{0}\frac{\sigma }{1+i\sigma \frac{%
\pi ^{2}}{2R\omega }},
\end{equation*}%
where $\sigma $ is a 2D conductivity of the layer including relaxation
processes. Using Eq.~(\ref{31}) for the magnetic moment, we arrive at the
expression which generalizes Eq.~(\ref{diss}):%
\begin{equation}
m_{z}=m_{z}^{\left( 0\right) }\frac{\omega ^{4}}{\left( \omega ^{2}-\omega
_{p}^{2}\right) ^{2}+\omega ^{2}\tau ^{-2}},
\end{equation}
where $m_{z}^{\left( 0\right) }$ is the magnitude of the magnetic moment
generated by a circularly polarized field without including dissipation and
depolarization effects, $\omega _{p}=\sqrt{\frac{\pi ge^{2}p_{F}v_{F}}{%
8\hbar ^{2}R}}$ , where $\frac{\partial W}{\partial p}=v_{F}$. The resonant
frequency $\omega _{p}$ in Eq.~(A4) coincides up to a numerical factor with
the frequency of 2D plasmons in graphene at wavelength $2R$ ; see e.g. \cite%
{hwang2007}. In the limit $R\rightarrow \infty $ Eq.~(A4) gives the result
for an infinite medium.

\section{IFE in graphene beyond small perturbation}

Here we consider an incident radiation of an arbitrarily strong intensity
and go beyond the linear approximation. Let's again assume a circularly
polarized field given by Eqs.~(A1). The kinetic equation Eq.~(6) with $\mathbf{%
H}=\mathbf{0}$ and relaxation operator $\hat{Q}\left( f\right) =\frac{f_{F}-f%
}{\tau }$ takes the form%
\begin{equation}
\frac{\partial f\left( \mathbf{p},t\right) }{\partial t}-eE_{0}\cos \left(
\omega t\right) \frac{\partial f\left( \mathbf{p},t\right) }{\partial p_{x}}%
+eE_{0}\sin \left( \omega t\right) \frac{\partial f\left( \mathbf{p}%
,t\right) }{\partial p_{y}}=\frac{f_{F}\left( p\right) -f\left( \mathbf{p}%
,t\right) }{\tau }.
\end{equation}%
Its solution in quadratures can be found by the method of characteristics.
At times $t\gg \tau $ for any initial conditions the solution approaches
\begin{equation}
f=e^{-\frac{t}{\tau }}\frac{1}{\tau }\int_{0}^{t}dt^{\prime }e^{\frac{%
t^{\prime }}{\tau }}f_{F}\left[ p_{x}+\frac{eE_{0}}{\omega }\left( \sin
\omega t-\sin \omega t^{\prime }\right) ,p_{y}+\frac{eE_{0}}{\omega }\left(
\cos \omega t-\cos \omega t^{\prime }\right) \right]
\end{equation}%
After cumbersome but fairly straightforward derivation, the surface current
density $\mathbf{j=}-egv_{F}\int \frac{\mathbf{p}}{p}fd^{2}p$ can be found:%
\begin{equation}
j_{x}=-en_{F}V_{x}\left( t\right) ,\ \ j_{y}=-en_{F}V_{y}\left( t\right).
\end{equation}%
Here the functions $V_{x,y}\left( t\right) $ are given by%
\begin{equation}
V_{x}\left( t\right) =\frac{v_{F}}{1-e^{-\frac{2\pi }{\left\vert \omega
\right\vert \tau }}}\int_{0}^{\frac{2\pi }{\left\vert \omega \right\vert
\tau }}e^{-z}\Phi \left( \frac{eE_{0}}{\omega p_{F}},\omega \tau z\right)
\left\{ \left[ 1-\cos \left( \omega \tau z\right) \right] \sin \left( \omega
t\right) +\sin \left( \omega \tau z\right) \cos \left( \omega t\right)
\right\} dz
\end{equation}%
\begin{equation}
V_{y}\left( t\right) =\frac{v_{F}}{1-e^{-\frac{2\pi }{\left\vert \omega
\right\vert \tau }}}\int_{0}^{\frac{2\pi }{\left\vert \omega \right\vert
\tau }}e^{-z}\Phi \left( \frac{eE_{0}}{\omega p_{F}},\omega \tau z\right)
\left\{ \left[ 1-\cos \left( \omega \tau z\right) \right] \cos \left( \omega
t\right) +\sin \left( \omega \tau z\right) \sin \left( \omega t\right)
\right\} dz
\end{equation}%
where%
\begin{equation}
\Phi \left( \frac{eE_{0}}{\omega p_{F}},\omega \tau z\right) =\left( \frac{%
2eE_{0}}{\pi \omega p_{F}}\right) \int_{0}^{\pi }\frac{\sin ^{2}\alpha }{%
\sqrt{1+4\left( \frac{eE_{0}}{\omega p_{F}}\right) ^{2}\sin ^{2}\left( \frac{%
\omega \tau z}{2}\right) +4\left\vert \frac{eE_{0}}{\omega p_{F}}\sin \left(
\frac{\omega \tau z}{2}\right) \right\vert \cos \alpha }}d\alpha .
\end{equation}%
It follows from (B3-B6) that the surface current density vector can be
presented in the form of Eqs.~(\ref{a2}), in which%
\begin{equation}
j_{0}=ev_{F}n_{F}F\left( \frac{eE_{0}}{\omega p_{F}},\omega \tau \right) ,
\end{equation}

\begin{equation}
\begin{array}{l} 
F\left( \frac{eE_{0}}{\omega p_{F}},\omega \tau \right) =\left( 1-e^{-\frac{
2\pi }{\left\vert \omega \right\vert \tau }}\right)^{-1} \times \nonumber  \\
  \left( \left\{
\int_{0}^{\frac{2\pi }{\left\vert \omega \right\vert \tau }}e^{-z}\Phi
\left( \frac{eE_{0}}{\omega p_{F}},\omega \tau z\right) \left[ 1-\cos \left(
\omega \tau z\right) \right] dz\right\} ^{2}  + \left\{
\int_{0}^{\frac{2\pi }{\left\vert \omega \right\vert \tau }}e^{-z}\Phi
\left( \frac{eE_{0}}{\omega p_{F}},\omega \tau z\right) \sin \left(
\omega \tau z\right)  dz\right\} ^{2}   \right)^{1/2}.
\end{array}
\end{equation} 
The value of the phase shift $\phi $ does not matter in this case. 

\begin{figure}[htb]
\begin{center}
\includegraphics[scale=0.6]{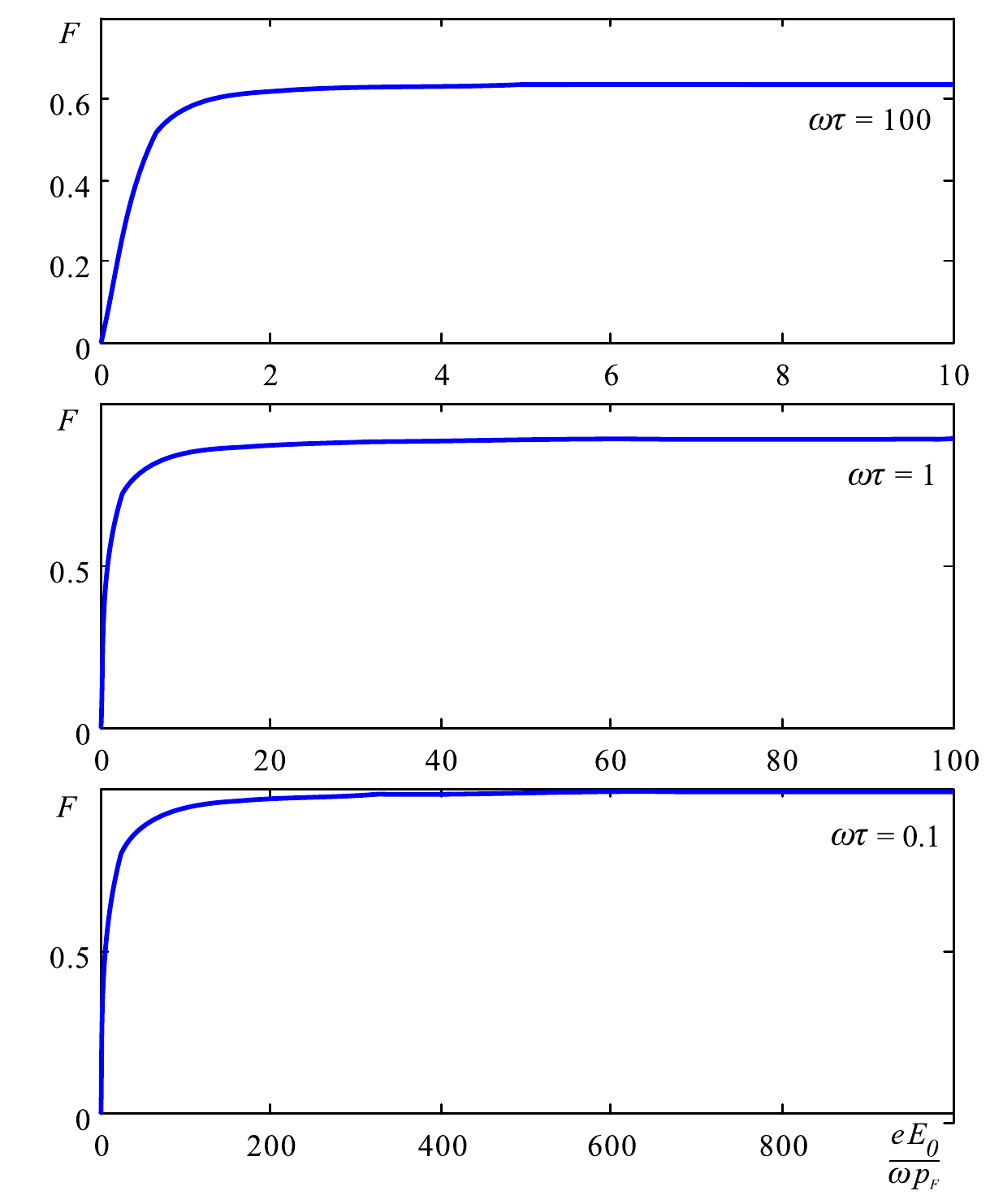}
\caption{ $F\left( \frac{eE_{0}}{\omega p_{F}},\omega
\tau \right) $ as a function of the parameter $\frac{eE_{0}}{\omega p_{F}}$ at different $
\omega \tau $. }
\label{fig5}
\end{center}
\end{figure}

Figure 5 shows the dependence $F\left( \frac{eE_{0}}{\omega p_{F}},\omega
\tau \right) $ on the parameter $\frac{eE_{0}}{\omega p_{F}}$ at different $%
\omega \tau $. There is an obvious saturation effect at $\frac{eE_{0}}{\omega
p_{F}}\gg 1$.

The current defined by Eqs.~(\ref{a2}),(B7) corresponds to the surface polarization
given by Eq.~(A3). Using the expression Eq.~(\ref{31}) for the magnetization, we
arrive at
\begin{equation}
\label{nonlin}
m_{z}=\frac{en_{F}v_{F}^{2}}{2c\omega }F^{2}\left( \frac{eE_{0}}{\omega p_{F}%
},\omega \tau \right) .
\end{equation}%
For weak fields, when $\frac{eE_{0}}{\omega p_{F}}\ll 1$, we have the limit%
\begin{equation*}
\Phi \left( \frac{eE_{0}}{\omega p_{F}},\omega \tau z\right) \cong \frac{%
eE_{0}}{\omega p_{F}},\ \ \ \ \ F\left( \frac{eE_{0}}{\omega p_{F}},\omega
\tau \right) \cong \frac{eE_{0}}{p_{F}\sqrt{\tau ^{-2}+\omega ^{2}}}
\end{equation*}%
In this case Eq.~(\ref{nonlin}) is reduced to Eq.~(\ref{diss}) for $\tilde{E}_{y}=-i$ $\tilde{E}%
_{x}$, $\tilde{E}_{x}=E_{0}$ .

The expression in Eq.~(\ref{nonlin}) allows one to estimate the magnitude of the IFE
for strong fields, when $\frac{eE_{0}}{\omega p_{F}}\geq 1.$\ \


\end{document}